\magnification=1120
\parskip=5.3pt 
\parindent=25pt
\baselineskip=17.5pt
\input amssym.def
\input amssym
\line{\hfil{DAMTP-R/98/15}}
\line{\hfil{Report 98-5~~~~~~~~}}
\line{\hfil{hep-th/9803093~~~}}
\line{\hfil{March, 1998~~~~~~~}}
\topglue .5in
\pageno=0
\footline={\ifnum \pageno <1 \else \hss \folio \hss\fi}
\vskip .55in
\centerline{\bf SUMMABILITY OF SUPERSTRING THEORY}
\vskip .5in
\centerline{Simon Davis}
\vskip .4in
\centerline{Department of Applied Mathematics and Theoretical Physics}
\vskip 5pt
\centerline{University of Cambridge}
\vskip 5pt
\centerline{Silver Street, Cambridge CB3 9EW, England} 
\vskip 5pt
\centerline{and}
\vskip 5pt
\centerline{School of Mathematics and Statistics}
\vskip 5pt
\centerline{University of Sydney}
\vskip 5pt
\centerline{NSW 2006, Australia}
\vskip .35in
\noindent{\bf Abstract.} Several arguments are given for the summability
of the superstring perturbation series.  Whereas the Schottky group 
coordinatization of moduli space may be used to provide refined estimates of 
large-order bosonic string amplitudes, the super-Schottky group variables 
define a measure for the supermoduli space integral which leads to upper
bounds on superstring scattering amplitudes.
\vskip .4in
\noindent
{\bf PACS numbers}: 0380, 0460, 1117
\hfil\break
{\bf Keywords}: superstring, moduli, genus, partitions, amplitudes, exponential

\vfill
\eject

The genus-dependence of superstring scattering amplitudes has been estimated 
recently using the super-Schottky coordinatization of supermoduli space.  The 
$N$-point $g$-loop amplitudes in Type IIB superstring theory have 
been found to grow exponentially with the genus, 
$~(4\pi(g-1))^{N-1}$  
$~\cdot~f(B_K, B_K^\prime, B_H, B_H^\prime, B_B, B_B^\prime)^{3g-3}~$, 
where $B_K,~B_K^\prime,~B_H,~B_H^\prime,~B_B$ and $B_B^\prime$ are the bounds
of integrals over Schottky group parameters which represent 
degenerating and non-degenerating moduli respectively [1].  As this 
genus-dependence differs significantly from the large-order growth of 
field theory amplitudes, several  arguments in support of the conclusion shall 
be put forward.

The advantage of using the Schottky parametrization of moduli space in the 
study of the growth of integrals representing the scattering amplitudes is 
that the dependence of these integrals on the genus is directly 
linked to the limits for each of the Schottky group variables.  By 
introducing a genus-independent cut-off on the length of closed geodesics,
in the string worldsheet metric, to regulate infrared divergences [2],
and then translating this condition to restrict the integration region
in the fundamental domain of the modular group, it can be shown that  
the sources of infrared and large-order divergences are identical, as
they both arise, in particular, from the genus-dependence of the
$\vert K_n\vert$ limit, $\vert K_n\vert~\sim~{1\over g}$ [3][4]. 
For the superstring, this cut-off is no longer necessary, and the entire 
fundamental region is required for the supermoduli space integrals.  
The introduction of supersymmetry eliminates the infrared divergences 
because of the absence of the tachyon in the superstring spectrum, and 
therefore, the large-order divergences are eliminated simultaneously.  This 
is a consequence of the tachyon being the source of the divergences, rather 
than the instanton, which could remain in the theory even after the 
introduction of supersymmetry.

The validity of estimates based on the super-Schottky group measure depends 
on the range of the integration.  The use of the super-Schottky group 
measure [5] 
$$\eqalign{{1\over {d {\cal V}_{ABC}}}~\prod_{n=1}^g~&{{dK_n}\over {K_n^{3\over
2}}}
{}~{{dZ_{1n} dZ_{2n}}\over {Z_{1n}-Z_{2n}}} \left({{1-K_n}\over {1-
(-1)^{B_n} K_n^{1\over
2}}}\right)^2~[~det(Im~{\cal T})~]^{-5}
\cr
&\cdot\prod_\alpha~^\prime \prod_{p=1}^\infty \left({{1-(-1)^{N_\alpha^B}
K_\alpha^{p-{1\over
2}}}
\over {1-K_\alpha^p}} \right)^{10}~\prod_\alpha~^\prime \prod_{p=2}^\infty~
\left({{1-K_\alpha^p}
\over {1-(-1)^{N_\alpha^B} K_\alpha^{p-{1\over 2}}}}\right)^2
\cr}
\eqno(1)$$
with infinitesimal super-projective invariant volume element 
$$\eqalign{d {\cal V}_{ABC}~&=~{{dZ_A dZ_B dZ_C}\over
{[(Z_A-Z_B)(Z_C-Z_A)(Z_B-Z_C)]^{1\over 2}}}\cdot {1\over {d\Theta_{ABC}}}
\cr
\Theta_{ABC}~&=~{{\theta_A(Z_B-Z_C)+
\theta_B(Z_C-Z_A)+\theta_C(Z_A-Z_B)+\theta_A
\theta_B\theta_C}\over {[(Z_A-Z_B)(Z_C-Z_A)(Z_B-Z_C)]^{1\over 2}}}
\cr}
\eqno(2)$$
and with super-period matrix 
$${\cal T}_{mn}~=~{1\over {2 \pi i}} \left[~ln~K_n~\delta_{mn}~+~
\sum_\alpha~^{(m,n)}~ln \left[{{Z_{1m}-V_\alpha Z_{1n}}\over {Z_{1m}-V_\alpha
Z_{2n}}} {{Z_{2m}- V_\alpha Z_{2n}}\over {Z_{2m}-V_\alpha Z_{1n}}} \right]
\right]
\eqno(3)$$
defines a splitting of a subset of supermoduli space, since the separation
of even and odd Grassmann coordinates is maintained over the entire region, a  
necessary condition for integration over the odd moduli.  Since the global
obstruction to the splitting of supermoduli space can be circumvented by 
removing a divisor $D_g$ of codimension greater than or equal to one from the
stable compactification of supermoduli space $s{\bar {\cal M}}_g$ [6], the 
integral over all of supermoduli space in superstring scattering amplitudes 
differ in the large-genus limit, therefore, from the estimates based on these 
subdomains by a contribution from the divisor, given by tadpoles of massless 
physical states at lower genera.  Assuming an exponential dependence 
up to genus $g-1$, and multiplying this bound by the number of degeneration 
limits corresponding to the splitting of the super-Riemann surface into two 
components, an estimate of the contribution of the divisor can be made.  
Eliminating the vanishing of two of the B-cycles by $OSp(2\vert 1)$ 
invariance of the super-Schottky uniformization, so that 
the contribution of the divisor will be bounded by   
$$(2g-2)~\cdot~(4\pi~(g-2))^{N-1}~
f(B_K,B_K^\prime,B_H,B_H^\prime,B_B,B_B^\prime)^{3g-6}~+
~2~\sum_{i=1}^{[{g\over 2}]}~d_i~A_i~A_{g-i}
\eqno(4)$$
where $A_i$ and $A_{g-i}$ represent amplitudes for the $i$ and $g-i$
components and $d_i$ is determined by the string propagator, the length
of the connecting tube and the moduli described by coordinates around the
punctures of the pinched surface.  The two terms in equation (4) represent the 
contributions of two different sets of divisors $\Delta_0$ and $\Delta_i$.

The boundary of supermoduli space can also be approached by considering
the degeneration of cycles non-homologous to zero.  Restricting to the
complex-valued part of the super-Riemann surface, of genus $g$,  the 
3$g$~-~3 degeneration limits are associated with the 
pinching of  A-cycles, B-cycles, which immediately can be identified
with the divisor $\Delta_0$, and C-cycles.  The C-cycles can be interchanged 
with the dividing cycles in a decomposition of moduli space based 
on the length and twist
parameters of 3$g$-3 non-intersecting closed geodesics on the surface.
Consequently, the counting of different partitions of the surface, which 
shall be required for an estimate of the total
amplitude, is equivalent when the remaining degeneration limits are viewed as
the vanishing of C-cycles or dividing cycles.  
 
Finiteness of the superstring amplitudes in each of the degeneration limits
$\vert K_n\vert \to 0$, 
\hfil\break
$n~=~1,...,~g$, $\vert H_m\vert \to 0$, 
$m~=~2,...,~g-1$, $\vert B_m\vert \to 0,~m~=~2,...,~g$, implies that there are
bounds associated with the corresponding multiplier integrals $B_K$, $B_H$,
$B_B$ for each value of $n$ and $m$, and it will be demonstrated that
additional integrals associated with non-degenerating moduli are bounded by 
$B_K^\prime$, $B_H^\prime$ and $B_B^\prime$.

A generic Riemann surface will have $l_K$ non-degenerating A-cycles, 
$g - l_K$ degenerating A-cycles, $l_H$ non-degenerating B-cycles, two
B-cycles fixed by an SL(2,{\Bbb C}) transformation, $g -2 - l_H$ degenerating 
B-cycles, $l_B$ non-degenerating C-cycles and $g - 1 - l_B$ non-degenerating 
C-cycles.  Given that the integrals over the non-degenerating moduli
are less than $B_K^\prime$, $B_H^\prime$ and $B_B^\prime$, upper bound 
including all degeneration limits is then given 
by
$$\eqalign{&\sum_{l=0}^{3g-3}~\sum_{{Part.\{l_K,l_H,l_B\}} 
\atop {l_K+l_H+l_B=l}}\left({g\atop {l_K}}\right) 
\left({{g-2}\atop {l_H}}\right) 
\left({{g-1}\atop {l_B}}\right)~B_K^{g-l_K} B_K^{\prime l_K}
B_H^{g-2-l_H} B_H^{\prime l_H} B_B^{g-1-l_B} B_B^{\prime l_B}
\cr
&~~~=~{{g!(g-2)!(g-1)!}\over {(3g-3)!}}~\sum_{l=0}^{3g-3}~
\sum_{{Part.\{l_K,l_H,l_B\}}\atop {l_K+l_H+l_B=l}}
~{{(3g-3)!}\over {l_K!(g-l_K)!l_H!(g-2-l_H)! l_B!(g-1-l_B)!}}
\cr  
&~~~~~~~~~~~~~~~~~~~~~~~~~~~~~~~~~~~~~~~~~~~~~~~~~~~~~~ \cdot B_K^{g-l_K}
B_K^{\prime l_K} B_H^{g-l_H} B_B^{g-l_B} B^{\prime l_B}
\cr}
\eqno(5)
$$
Since the last sum is bounded by
$$\sum_{{l_1,...,l_6}\atop {l_1+...+l_6=3g-3}}
~\left({{3g-3}\atop {l_1 l_2 l_3 l_4 l_5 l_6}}\right) B_K^{l_1} 
B_K^{\prime l_2} B_H^{l_3} B_H^{\prime l_4} B_B^{l_5} B_B^{\prime l_6}
~=~(B_K+B_K^\prime+B_H+B_H^\prime+B_B+B_B^\prime)^{3g-3}
\eqno(6)
$$
the entire expression is less than
$${{g!(g-2)!(g-1)!}\over {(3g-3)!}}~(B_K+B_K^\prime+B_H+B_H^\prime
+B_B+B_B^\prime)^{3g-3}
\eqno(7)
$$
If the ratio ${{B_K+B_K^\prime+B_H+B_H^\prime+B_B+B_B^\prime}
\over {min(B_K,B_H,B_K)}}$ is denoted by ${\tilde K}$, then
$${{(B_K+B_K^\prime+B_H+B_H^\prime+B_B+B_B^\prime)^{3g-3}}\over 
{B_K^g B_H^{g-2} B_B^{g-1}}}< {\tilde K}^{3g-3}
\eqno(8)
$$
and the N-point scattering amplitude is bounded by
$c_1 c_2^g (4 \pi (g-1))^N B_K^g B_H^{g-2} B_B^{g-1}$ if $c_1=2$
and $c_2~=~{{\tilde K}\over 3}$.

Contributions of degenerate Riemann surfaces to the amplitude  
can be compared with integrals over the region of parameter space defined by
the condition of genus-independent bounds for the multipliers and 
distances between the fixed points.  For the latter category of surfaces,
the uniformizing super-Schottky groups have multipliers and fixed points which
satisfy the inequalities
$~\epsilon_0 \le \vert K_n \vert \le \epsilon_0^\prime~$, $~\delta_0 \le
\vert \xi_{1n} - \xi_{2n} \vert \le \delta_0^\prime~$.
Integration over this range produces the following expression

$$\eqalign{\vert~v(\xi_{11}^0,\xi_{21}^0,\xi_{1g}^0,\theta_{11}^0,
\theta_{1g}^0)\vert^2~{{\pi^{2g-3}}\over {4^{2g-2}}}
&\left[\left(ln~{1\over {\epsilon_0^\prime}}\right)^{-4}
~-~\left(ln~{1\over {\epsilon_0}}\right)^{-4}\right]^g
\left[{1\over {\delta_0^2}}-{1\over {\delta_0^{\prime 2}}}\right]^{g-2}
\cr
&(\delta_1^{\prime 2}~-~\delta_1^2)^{g-2} 
\left[1~-~{{(\delta_0+\delta_0^\prime)^2}
\over {4 \delta_1^{\prime 2}} }\right]^{g-2}
\cr
&\cdot [~contribution~from~the~B_m~multipliers~]
\cr}
\eqno(9)
$$
since there is a projective mapping of the isometric circles to 
configurations with $\delta_1 < \vert \xi_{1m}\vert < \delta_1^\prime$ and 
the $\vartheta$ integrations can be absorbed in

$$\prod_{i=2}^{g-1}~\int~d\vartheta_{1i} \vartheta_{1i}
~\prod_{j=1}^g~\int~d\vartheta_{2j} \vartheta_{2j}
\eqno(10)
$$
In equation (9), the first factor arises from the Jacobian factor resulting
from the residual $OSp(2\vert 1)$ symmetry of the super-Schottky
parameterization used to select the locations of three of the 
fixed points $\xi_{11}^0$, $\xi_{21}^0$ and $\xi_{1g}^0$ and two of the 
Grassmann variables $\theta_{11}^0$ and $\theta_{1g}^0$, and the final 
factor on the second line represents the reduction of the integral 
as a result of requiring non-overlapping of the isometric disks in the 
region of the complex plane occupied by the disks.  From equation (2),
it follows that
$$\eqalign{{1\over {d{\cal V}_{ABC}}}~&=~
{1\over {d\xi_A d\xi_B d\xi_C d\theta_B
d\theta_C}}\left[Z_B~-~Z_C~-\theta_B \theta_C~-~{1\over 2} {{\theta_B}
\over {Z_A~-~Z_B}}~+~{1\over 2} {{\theta_C}\over {Z_C~-~Z_A}}\right]
\cr
~&~~~~+~\left[\theta_C~-~{1\over 2} {1\over {Z_A~-~Z_B}} \right] 
~{{(d\xi_A~-~d\xi_B)}\over {d\xi_A d\xi_B d\xi_C d\theta_A
d \theta_B d\theta_C}}
\cr
~&~~~~+~cyclic~permutations~of~(A,B,C)
\cr}
\eqno(11)
$$
The leading term in the expansion with respect to the Grassmann variables,
after excluding terms involving integration over a set of parameters
other than $(3g-3)$ fixed points and multipliers and $(2g-2)$ odd moduli, is
$$\left[{{(\xi_{11}^0~-~\xi_{21}^0)}\over {d\xi_{11}d\xi_{21} d\xi_{1g}
d\theta_{11} d\theta_{21}}}~+~{{(\xi_{1g}^0~-~\xi_{11}^0)}
\over {d \xi_{11} d\xi_{21} d\xi_{1g} d\theta_{11} d\theta_{1g}}}
~+~{{(\xi_{21}^0~-~\xi_{1g}^0)}\over {d\xi_{11} d\xi_{21} d\xi_{1g} 
d\theta_{21} d\theta_{1g}}}\right]~\times~c.c.
\eqno(12)
$$
The functional dependence of $\vert v(\xi_{11}^0,\xi_{21}^0,\xi_{1g}^0, 
\theta_{11}^0, \theta_{1g}^0)\vert^2$
can be deduced from (12) after relabelling the Grassmann variables to 
remove the repetition  arising from cyclic permutations.    
The $B_m$ part of the measure is

$$\prod_{m=2}^g {{d B_m}\over {B_m^{3\over 2}} }\cdot~\prod_{m=2}^g
~{{d{\bar B}_m}\over {{{\bar B}_m}^{3\over 2}}  }
~=~\prod_{m=2}^g~{{\vert B_m\vert~d\vert B_m \vert~d \theta_m^B}
\over {\vert B_m\vert^3}}
\eqno(13)
$$
Since $\vert \xi_{2m}\vert~=~\prod_{j=2}^m~\vert B_j \vert $, it follows that
if $\delta_2 < \vert \xi_{2n} \vert < \delta_2^\prime$, the range for
$\vert B_j \vert$ can be chosen to be $[\delta_2^{1\over g},
\delta_2^{\prime {1\over g}}]$.  Because 
$$lim_{g\to \infty}\left(1+{{ln~\delta_2}\over g}\right)
~=~lim_{g\to\infty}~\delta_2^{{1\over g}}
\eqno(14)
$$
the magnitude of the integral depends on the choice of
$\delta_2$.  If $\delta_2$ is fixed to be a finite number greater than zero,
then the lower limit of $\vert B_j\vert$ for large $g$, is 
$1+{{ln~\delta_2}\over g}$, and the integral is
$$\prod_{m=2}^g~\left[{{ln~\delta_2^\prime~-~ln~\delta_2}\over g}
~+~{{ln~\delta_2^2~-~ln~\delta_2^{\prime 2}}\over {g^2}}~+~O(g^{-3})
~\right]
\eqno(15)
$$
which, of course, significantly decreases the overall genus-dependence of
the integral.  However, it is clear that the integration region can be
enlarged so that the lower limit for $\vert B_j\vert$ is ${\tilde \epsilon}$.
The lower limit for $\vert \xi_{2m} \vert$ then would be 
${\tilde \epsilon}^{m-1}$, which tends to zero as $m\to g$.  The 
$\vert B_m\vert$ integrals are therefore
$$\prod_{m=2}^g~\int_{\tilde \epsilon}^{1+{{ln~\delta_2^\prime}\over g}}
~{{d\vert B_m\vert}\over {\vert B_m\vert^2}}~=~\prod_{m=2}^g~
\left[{{-1}\over {\vert B_m\vert}}
\bigg\vert_{\tilde\epsilon}^{1+{ln~{{\delta_2^\prime}\over g}}}\right]
~=~\prod_{m=2}^g~\left[{1\over {\tilde \epsilon}}~-~
{1\over {1~+~{{ln~\delta_2^\prime}\over g}}}\right]
~<~{1\over {{\tilde \epsilon}^{g-1}}}
\eqno(16)
$$
The contribution from the $B_m$ multipliers is bounded by ${{(2\pi)^{g-1}}
\over {{\tilde \epsilon}^{g-1}}}$ and therefore it does not
affect the overall exponential dependence of the integral over this region
in the parameter space.

The limits $\vert B_m\vert \to 0$ represent the degeneration of C-cycles
since $\vert \xi_{2j}\vert \to 0$, $j~=~m,...,~g$.  The Riemann surface
splits into two components of genus $m-1$ and $g-m+1$ and each
component appears to be a point from the perspective of the other 
component [7].  

Since modular transformations map A-cycles, B-cycles and C-cycles into
each other, genus-dependent lower limits for the ranges of the
variables $\{\vert \xi_{2m}\vert\}$ would take the form 
${{\delta_2}\over {g^{2{\hat q}}}}$.  The lower
bound for $\vert B_j\vert$ would then be $\left({{\delta_2}
\over {g^{2{\hat q}}}}\right)^{1\over g}$.  Since $g^{1\over g} \to 1$ as 
$g \to \infty$, it follows that the lower limit of the $\vert B_m\vert$ 
integral tends to 1.  In this case, the  value of ${\tilde \epsilon}$ can be
chosen to be any constant less than 1. 

In the degeneration limit $\vert B_m\vert \to 0$, the superstring amplitude
is finite, so that the integral of the entire supersymmetric measure over
the range $\left[0,\left({{\delta_2}\over {g^{2{\hat q}} } }\right)^{1\over g}
\right]$ will be bounded.  Since the $B_m$ integrals over the
neighbourhood of the boundary and the interior of supermoduli space 
are bounded by exponential functions of the genus, their sum will possess
the same property.                        

Given a holomorphic slice on the subset of supermoduli space, 
$s{\cal M}_g~-~{\cal N}(D)_g$, with ${\cal N}(D_g)$ being a neighbourhood of 
the divisor, an analytic transformation to the super-Schottky group variables
maps this integration region to a subdomain of the fundamental region
in the super-Schottky parameter space, excluding a neighbourhood
of the boundary.  If the separation between the neighbourhood 
of the compactification divisor and the interior region of moduli space 
were to give rise to genus-dependence in the ranges of the 
super-Schottky variables, computations of the multiplier integrals 
with genus-dependent limits imply that they still should have 
finite upper bounds.  This can be seen, for example, 
by considering the integral 
$$\int~{{d\vert K_n\vert}\over {\vert K_n\vert~\left(ln~\left({1\over
{\vert K_n\vert}}\right)\right)^5}}
~=~{1\over 4}\left[ln~\left({1\over {\vert K_n\vert}}\right)\right]^{-4}
\eqno(17)
$$
with mixed integration limits such as $\left[{{\epsilon_0}\over 
{g^{1-2q^\prime}}}, \epsilon_0^\prime \right]$, which gives the result
$$-{1\over 4}[(1-2 q^\prime)~ln~g~-~ln~\epsilon_0]^{-4}
~+~{1\over 4}\left[ln~\left({1\over {\epsilon_0^\prime}}\right)\right]^{-4}
\eqno(18)
$$
Moreover, as long as $\epsilon_0^\prime$ does not asymptotically approach 1
at a sufficiently fast rate as $g \to \infty$, equation (18) implies that
the upper bound for the integral will be an exponential function of the
genus.  This can be determined by explicit calculations of the location of
the fundamental region of the modular group in Schottky parameter
space.  At genus 1, the fundamental region of $SL(2;{\Bbb Z})$ in the
upper half-plane is well-known, and the maximum value of $\vert K\vert
~=~\vert e^{2\pi i \tau}\vert~=~e^{-2\pi~Im~\tau}$ occurs at the minimum
value of $Im~\tau$, ${{\sqrt 3}\over 2}$, and equals $e^{{-\sqrt 3}\pi}$.
At higher genus, the determinant condition can be used to show that
$(Re~\tau)_{nn}^2~+~(Im~\tau)_{nn}^2~\ge~1$, $n~=~1,...,~g$ [8].  Since
one of the conditions defining the fundamental region of the symplectic
modular group is $-{1\over 2}~\le~(Re~\tau)_{mn}~\le~{1\over 2}$, the
squares of the diagonal elements of $(Re~\tau)$ are restricted to the
interval $[~0, {1\over 4}~]$ and $(Im~\tau)_{nn}^2~\ge~{3\over 4}$.
Positive definiteness of imaginary part of the period matrix implies that
the positive root should be chosen, $(Im~\tau)_{nn}~\ge~{{\sqrt 3}\over 2}$.  

The following argument also may be useful in setting upper limits for the
range of $\vert K_n\vert$.  One of the conditions defining the fundamental 
region of the modular group is $\vert~det(C\tau + D)~\vert~\ge~1$ for
$\left(\matrix{A&B&
        \cr
       C& D&
        \cr}\right) \in Sp(2g; {\Bbb Z})$.  Since 
$$\vert~det(C\tau + D)~\vert~=~\vert~det (Im~\tau)\vert 
                             ~\vert~det(C~-~iC(Re~\tau)(Im~\tau)^{-1}
                                     ~-~i D(Im~\tau)^{-1})~\vert
\eqno(19)
$$
the determinant will be greater than one for all $C$, $D$ when
$\vert~det(Im~\tau)~\vert>~b~> 1$ for some number $b$, so that
$\vert~det(C~-~iC(Re~\tau)(Im~\tau)^{-1}~-~iD(Im~\tau)^{-1})~\vert$
will be bounded below when $det~C~\ne~0$ and equal to
$\vert~det~D~\vert~\vert~det(Im~\tau)^{-1}~\vert~\ge~\vert~
det(Im~\tau)~\vert^{-1}$  when $C~=~0$.  Moreover, the minimum value of 
${{tr(Im~\tau)}\over g}$ is
$$ln{1\over {\epsilon_0^\prime}}~+~{1\over g}~\sum_{n=1}^g~l_{nn}
\eqno(20)
$$
where $l_{nn}$ is the greatest lower bound for 
$$\sum_\alpha~^{(n,n)}~ln~\left\vert~{{\xi_{1n}~-~V_\alpha~\xi_{2n}}
\over {\xi_{1n}~-~V_\alpha \xi_{1n}}}
{{\xi_{2n}-V_\alpha \xi_{1n}}\over {\xi_{2n}~-~V_\alpha \xi_{2n}}} 
\right\vert
\eqno(21)
$$
If $\epsilon_0^\prime$ satisfies the inequality,
$$ln~{1\over {\epsilon_0^\prime}}~\ge~b^{1\over g}~-~{1\over g}~
\sum_{n=1}^g~l_{nn}
\eqno(22)
$$
the Schottky group multipliers and fixed points will lie in the
fundamental region of the symplectic modular group.  It has been
discussed previously how exponentials of the sums in (21) can
be estimated for different configurations of isometric circles [4].
Moreover, positive-definite symmetric real matrices
can be expressed as ${\cal Q}{\cal P}{\cal Q}^T$, where ${\cal P}$ diagonal
matrix with entries $p_k~>~0$ and ${\cal Q}$ is a triangular matrix 
$(q_{kl}),~q_{kl}~=~0,~k~>~l$ and $q_{kk}~=~1$, and the fundamental domain of 
the symplectic modular group is contained in the region defined by the 
inequalities $0~\le~p_k~\le~t\cdot p_{k+1}$ and $-t~\le~q_{kl}~\le~t$ [8].
These constraints can be applied to $(Im~\tau)$, with the value of $t$ being
related to the bound on the sum (21).
       
Both fermionic and bosonic contributions to one-loop superstring amplitudes 
can both be positive, since there are no odd moduli parameters 
at this genus, $det~(Im~\tau)~\ge~0$, and the measure 
is positive-definite.
The four-point one-loop amplitude contains an integral over the entire 
Riemann surface, so that
$s$-channel, $t$-channel and $u$-channel diagrams are included in the field 
theory limit of the string
diagrams.  The square of the absolute value of a holomorphic function, 
$\vert f(w)\vert^{16}$,
arising for fermions, is cancelled by a factor $[f(w)]^{-8}$ for 
right-moving modes $\alpha_n^i$ and
the complex conjugate factor $[f({\bar w})]^{-8}$ for left-moving modes 
${\tilde {\alpha}}_n^i$
in Type II superstring theories [9], eliminating a potential divergence in 
the integral over the modular parameter.

The positive fermionic and bosonic contributions to the one-loop amplitudes 
imply that the
superstring amplitudes could receive contributions growing at nearly 
identical rates with
respect to the genus, thus maintaining the approximately factorial rate of 
growth of the bosonic string partition function.  However, this 
property is changed by the odd modular parameters at higher genus.  This 
implies that the determinant of the imaginary part of the super-period 
matrix is no longer necessarily positive-definite.  Even though the 
remaining part of the measure consists of the absolute square of a holomorphic 
function of the supermoduli parameters, the measure is not necessarily 
positive-definite and the additional contribution to the amplitudes 
at higher genus may not occur.  Indeed, supersymmetric theories 
generally exhibit better large-order behaviour, and this is confirmed 
by the growth of the superstring amplitudes.  

A recent general study of divergences in perturbation theory reveals that 
they are linked to the violation of the hypothesis of Lebesgue's Dominated
Convergence Theorem concerning the formal manipulation of interchanging
the sum and the integral to obtain the series in the functional integral
formalism [10].  This problem is resolved by introducing a functional
representation of the characteristic function which cuts off the region
in field space representing the source of the divergences.  Specifically, it
is demonstrated that there exists a convergent series expansion of the
partition function for $\lambda {{\phi^4}\over {4!}}$ theory such that the
partial sums $Z_N(\lambda)$ tend to the exact finite value for $\lambda > 0$.
It is also known that $Z(\lambda)$ is non-analytic at $\lambda=0$, based on
an argument similar to that applied to quantum electrodynamics.  Even
though the partial sums $Z_N(\lambda)$ satisfy the hypothesis of the
Dominated Convergence Theorem, the non-analyticity of $Z(\lambda)$ at
$\lambda=0$ arises in the limit $N \to \infty$.  Techniques of this kind 
have already been used in the regulation of infrared divergences in closed 
bosonic string theory and the introduction of finite ultraviolet and infrared
cut-offs in QED with fermions yielding a convergent perturbation expansion 
[11].  Since a version of quantum electrodynamics and the non-abelian gauge 
theories of the standard model should arise in the low-energy limit of 
superstring theory, it is useful to have a framework in which non-analyticity 
in the coupling constant plane still can be derived from convergent or summable
perturbation expansions rather than series with terms increasing in
magnitude at a factorial rate with respect to the order.
This provides confirmation of the type of growth that has been obtained in 
the study of superstring scattering amplitudes.

It has been established that field theory amplitudes arise as limits of
string theory amplitudes by identifying different types of field theory
diagrams with corners of moduli space.  Amplitudes with two, three, four
and five gluons at one loop in SU(N) Yang-Mills theory, for example, have
been evaluated using certain open bosonic string amplitudes in the limit of
vanishing Regge slope $\alpha^\prime \to 0$ [12][13].  In general, the corner
of moduli space is defined by letting the complex string moduli space
approach the pinching limit so that they can be mapped to Schwinger proper
times.  The neighbourhood of the singular point at the boundary of moduli
space can be identified when the string worldsheet begins to resemble a
specific $\Phi^3$ diagram [13].  To define the pinching limit properly, it
is sufficient to cut open all of the loops of an $N$-point $g$-loop diagram
to form a 2$g$+$N$-point $\Phi^3$ tree diagram.  The tree consists of a main
branch and side branches, each with its own Schwinger proper time flow.  The
vertices of the diagram can be selected to be the points 
$\{z_1,...,z_N; \xi_{11},\xi_{21},...,\xi_{1n},\xi_{2n}\}$ and the
pinching limits are represented by the approach of the vertices towards the
branch endpoints $z_{B_i}$, $i = 1,..., N_b$, so that $\vert z - z_{B_i}\vert
\ll 1$ [14].  Many different labellings correspond to the same $\Phi^3$
diagram, including those obtained by interchanging $\xi_{1n}$ with
$\xi_{2n}$ or permuting the $g$ triplets $\{k_n,\xi_{1m},\xi_{2m}\}$.
Since the relevant $\Phi^3$ diagrams are identified with the corners
of moduli space which are neighbourhoods of components of the compactification
divisor ${\cal D}~=~{\cal D}_0 \cup {\cal D}_i$, it is sufficient to count
the degeneration limits, given by the partitions of $l$ with each of the
addends less than $g$, $g-2$ and $g-1$ and the sum less than or equal to
$3g-3$.  The upper bound on the number of relevant diagrams is then
$$\sum_{l=0}^{3g-3}~\sum_{Part.\{l_K,l_H,l_B\}}~\left({g\atop {l_K}}\right)
\left({{g-2}\atop {l_H}}\right)~\left({{g-1}\atop {l_B}}\right)
~<~3^{3g-3}~\left({{g-1}\over {g-2}}\right)^2
\eqno(23)
$$
The remainder of the string integral includes surfaces well away from the
degeneration limit and the contribution has been estimated to be an
exponential function of the genus.  There is no direct identification of this
region with field theory diagrams, and it is preferable to bound the
supermoduli space integral over the region using the estimates for 
ranges in the super-Schottky parameter space lying in the interior of
the fundamental domain of the modular group.

For a closed bosonic string, the lower bound for the regularized partition 
function increases at an approximately factorial rate with respect to $g$ 
[2][3][4][15].  These results are similar to the
calculations of the Euler characteristic of the once-punctured moduli
space [16][17][18][19], duplicated in the counting of Feynman diagrams in
random surface theory and matrix models [20][21].  However, it is apparent 
that the magnitudes of the moduli space and supermoduli space integrals 
depend essentially on the measure.  The change from 
$\int \prod_{n=1}^g {{d^2 K_n}\over {\vert K_n \vert^4
~\left[ln \left({1\over {\vert K_n\vert}}\right)\right]^{13}}}$
in bosonic string theory to $\int~\prod_{n=1}^g~{{d^2 K_n}\over 
{\vert K_n\vert^2~\left[ln \left({1\over {\vert K_n\vert}}\right)\right]^5}}$
in superstring theory, reflecting the removal of the tachyon singularity
in the Neveu-Schwarz sector and the absence of the tachyon in the 
Ramond sector, is sufficient to render the total amplitude an exponential
function of the genus in the latter case, whereas it causes the former integral
to increase as ${{g^{2g}}\over {(ln~g)^{13~g}}}$, when the range of 
$\vert K_n\vert$ is $\left[{{\epsilon_0}\over g}, 
{{\epsilon_0^\prime}\over g}\right]$.  Furthermore, the upper bound (23)
on the number of different degeneration limits, implies that the
partitioning of the surface into components only leads to the upper bound
being multiplied by an exponential factor, because the counting of relevant 
diagrams in a cell decomposition of supermoduli space is restricted to a 
neighbourhood of the compactification divisor. 

    Superstring elastic scattering in the large-$s$ fixed-$t$ limit has been 
studied, and the amplitude has been calculated using covariant 
loop sewing techniques [22][23][24], revealing that the leading term in the 
expansion in powers of $s^{-1}$ factors at $g$-loop order 
into a product of $g+1$ tree amplitudes,
multiplied by the expectation value of a factorized operator.  The eikonal 
approximation,
involving a resummation of these contributions, restores unitarity in the 
theory and gives
an exponential result.  Restriction to one copy of the fundamental region 
implies that a
factor of ${1\over {(g+1)!}}$ should be included in the integration, removing 
the factorial
dependence obtained by integrating over the multipliers and the variables 
$~\rho_n~=~-ln~\left[{{\xi_{1n}}\over {\xi_{2n}}}~
{{z_1-\xi_{2n}}\over {z_1-\xi_{1n}}}~\right]~$,              
$~\sigma_n~=~ln~\left({{1-\xi_{2n}}\over {1-\xi_{1n}}} \right)~$ [4][23][24].
These results confirm the generic estimates.

    The estimates of the superstring amplitudes are also consistent with the 
growth of the special scattering amplitudes obtained from topological 
field theory.  In particular, the $(g!)^2$ dependence of amplitudes 
with $2g-2$ graviphotons and 2 gravitons [25][26] may be verified by
combining the exponential bounds on the supermoduli space integrals 
with the factorial bounds on the vertex operator integrals [1].

The insertion of Dirichlet boundaries in string worldsheets has been 
used to reproduce 
power-law behaviour associated with point-like structure in QCD [27][28][29].
Summing over orientable Riemann surfaces with Dirichlet boundary conditions 
by associating
the moduli with the positions and strengths of electric charges [30], one 
obtains non-perturbative
amplitudes of order $e^{-{1\over \kappa_{str}}}~A^{conn.}$
which will only be well-defined if the amplitudes for the closed surfaces are 
also summable.  Summability of the superstring perturbation series
would be confirmed by the extension of the positive-energy theorem to
superstring vacua corresponding to supersymmetric background geometries
such as ${\Bbb R}^{10}$.  Thus, the exponential dependence of the 
closed-surface amplitudes and the description of non-perturbative effects
are compatible, since the latter can be regarded as additional contributions
to the superstring path integral, separate from the sum over closed
surfaces.    
The insertion of the Dirichlet boundary is associated with a different class 
of surfaces, which may be confirmed by considering the effect on the shift
to the vacuum.  Although there may be a small amplitude for
the non-perturbative instability of the initial string vacuum state, this
still could be consistent with the positive-energy theorem, which might be 
circumvented through the use of zero-norm boundary states.
\vskip 8pt
\centerline{\bf Acknowledgements}
\vskip 4pt
\parindent=25pt
I would like to thank Prof. M. Green for useful discussions on string
perturbation theory.  
\hfil\break
This research has been initiated 
in the Department of Applied Mathematics and Theoretical Physics at the 
University of Cambridge in May, 1996, and the hospitality of 
Dr G. W. Gibbons and Prof. S. W. Hawking is gratefully acknowledged.  
The encouragement of Dr H. C. Luckock, which allowed this work could be 
completed at the University of Sydney, is also much appreciated.
\vskip 3pt
\baselineskip=2pt
\centerline{\bf References}
\vskip 3pt
\item{[1]}  S. Davis, `Modular Invariance and the Finiteness of Superstring 
Theory' DAMTP-R/94/27, hep-th/9503231
\item{[2]}  D. Gross and V. Periwal, Phys. Rev. Lett. ${\underline{60}}$
(1988) 2105 - 2108
\item{[3]}  S. Davis, Class. Quantum Grav. ${\underline{7}}$ (1990)
1887 - 1893
\item{[4]}  S. Davis, Class. Quantum Grav. ${\underline{11}}$ (1994)
1185 - 1200 
\item{[5]}  A. Bellini, G. Cristofano, M. Fabbrichesi and K. Roland,
Nucl. Phys. ${\underline{B356}}$ (1991) 69 - 116
\item{[6]}  J. J. Atick, G. Moore and A. Sen, Nucl. Phys. ${\underline{B308}}$
(1988) 1 - 101
\item{[7]}  E. Martinec, Nucl. Phys. ${\underline{B281}}$ (1987) 157 - 210
\item{[8]}  C. L. Siegel, ${\underline{Symplectic~Geometry}}$ 
(New York: Academic Press, 1964)
\item{[9]}  M. B. Green, J. Schwarz and E. Witten, ${\underline{Superstring~
Theory}}$ Vol. 1 (Cambridge: Cambridge University Press, 1986)
\item{[10]} S. A. Pernice and G. Oleaga, Phys. Rev. ${\underline{D57}}$
(1998) 1144 - 1158
\item{[11]}  E. Caianello, Nuovo Cimento, ${\underline{3}}$ (1956) 223 - 225 
\item{[12]}  Z. Bern, L. Dixon and D. A. Kosower, Phys. Rev. Lett.
${\underline{70}}$ (1993) 2677 - 2680
\item{[13]}  P. Di Vecchia, L. Magnea, A. Lerda, R. Russo and R. Marotta,
Nucl. Phys. ${\underline{469}}$ (1996) 235 - 286
\item{[14]}  K. Roland and H.-T. Sato, Nucl. Phys. ${\underline{B480}}$
(1996) 99 - 124
\item{[15]}  S. Davis, J. Math. Phys. ${\underline{36}}$(2) (1995) 648 - 663
\item{[16]}  R. C. Penner, `The Moduli Space of Punctured Surfaces'
San Diego 1986, Proceedings,
\hfil\break
${\underline{Mathematical~Aspects~of String~Theory}}$, 313 - 340
\item{[17]}  J. Harer and D. Zagier, Invent. Math. ${\underline{85}}$
(1986) 457 - 485
\item{[18]}  R. C. Penner, Commun. Math. Phys. ${\underline{113}}$
(1987) 299 - 339
\item{[19]}  C. Itzykson and J.-B. Zuber, Commun. Math. Phys. 
${\underline{134}}$ (1990) 197 - 207
\item{[20]}  D. J. Gross and A. A. Migdal, Phys. Rev. Lett. ${\underline{64}}$
(1990) 127 - 130
\item{[21]}  D. Kutasov and N. Seiberg, PHys. Lett. ${\underline{B251}}$
(1990) 67 - 72
\item{[22]}  J. L. Petersen, J. R. Sidenius and A. K. Tollsten, Nucl. Phys.
${\underline{B317}}$ (1989) 109 - 146
\item{[23]}  G. Cristofano, M. Fabbrichesi and K. Roland, Phys. Lett.
${\underline{B236}}$ (1990) 159 - 164
\item{[24]}  G. Cristofano, M. Fabbrichesi and K. Roland, Phys. Lett.
${\underline{B244}}$ (1990) 397 - 402
\item{[25]}  I. Antoniadis, E. Gava, K. S. Narain and T. R. Taylor,
Nucl. Phys. ${\underline{B413}}$ (1994) 162 - 184
\item{[26]}  M. Bershadsky, S. Cecotti, H. Ooguri and C. Vafa, Commun.
Math. Phys. ${\underline{165}}$ (1994) 311 - 427
\item{[27]}  M. B. Green, Phys. Lett. ${\underline{B266}}$ (1991) 325 - 336
\item{[28]}  M. B. Green, Phys. Lett. ${\underline{B302}}$ (1993) 29 - 37
\item{[29]}  M. B. Green, Phys. Lett. ${\underline{B329}}$ (1994) 435 - 443 
\item{[30]}  M. B. Green and J. Polchinski, Phys. Lett. ${\underline{B335}}$
(1994) 377 - 382

\end